\documentclass[%
aps,
prl,
reprint,
twocolumn,
superscriptaddress
]{revtex4-2}
\usepackage{amsmath}
\usepackage{mathtools} 
\usepackage{graphicx}
\usepackage{dcolumn}
\usepackage{bm}
\usepackage{hyperref}
\hypersetup{
	bookmarks=true,
	colorlinks=true,
	urlcolor=blue,
	citecolor=blue,
	linkcolor=blue, 
	pdftex,
	linktocpage=true, 
	linktoc=all,     
	hyperindex=true
}
\usepackage[mathlines]{lineno}

\usepackage{siunitx} 
\usepackage{comment} 
\usepackage{braket} 

\begin{document}


\title{Binary Dipolar Condensates of Dysprosium Isotopes with Tunable Spatial Order}
\author{Shenshuang Nie}
\thanks{These authors contributed equally to this work.}
\affiliation{Department of Physics and Guangdong Basic Research Center of Excellence for Quantum Science, Southern University of Science and Technology, Shenzhen 518055, China}%

\author{Zibin Jiang}%
\thanks{These authors contributed equally to this work.}
\affiliation{Department of Physics and Guangdong Basic Research Center of Excellence for Quantum Science, Southern University of Science and Technology, Shenzhen 518055, China}%

\author{Junrong Huang}%
\affiliation{Department of Physics and Guangdong Basic Research Center of Excellence for Quantum Science, Southern University of Science and Technology, Shenzhen 518055, China}%

\author{Xiao Luo}%
\affiliation{Department of Physics and Guangdong Basic Research Center of Excellence for Quantum Science, Southern University of Science and Technology, Shenzhen 518055, China}%

\author{Fucheng Qin}%
\affiliation{Department of Physics and Guangdong Basic Research Center of Excellence for Quantum Science, Southern University of Science and Technology, Shenzhen 518055, China}%

\author{Kaiyue Wang}%
\affiliation{Department of Physics and Guangdong Basic Research Center of Excellence for Quantum Science, Southern University of Science and Technology, Shenzhen 518055, China}%

\author{Mingyang Guo}%
\email{guomy@sustech.edu.cn}
\affiliation{Department of Physics and Guangdong Basic Research Center of Excellence for Quantum Science, Southern University of Science and Technology, Shenzhen 518055, China}%
\affiliation{State Key Laboratory of Quantum Functional Materials, Southern University of Science and Technology, Shenzhen 518055, China}%
\affiliation{Quantum Science Center of Guangdong-Hong Kong-Macao Greater Bay Area, Shenzhen 518045, China}

\date{\today}

\begin{abstract}

Dipolar quantum mixtures provide a route to many-body phases in which long-range anisotropic interactions couple with density, composition and spatial order. Here we realize a new quantum-degenerate dipolar mixture of $^{162}$Dy and $^{164}$Dy in a single-species-like apparatus. The mixture combines nearly matched single-particle Hamiltonians, tunable interactions and composition parameters, and isotope-resolved characterization. Tuning the interaction balance and relative composition reorganizes the coupled condensates from a miscible state into core--shell-like, side-by-side, and exchanged core--shell-like immiscible configurations. These results establish dysprosium isotope mixtures as a compact and versatile platform for multicomponent dipolar quantum matter, ranging from impurity physics to binary supersolidity.
	

\end{abstract}

\maketitle

Dipolar quantum gases of highly magnetic atoms have opened access to many-body physics beyond the contact-interaction paradigm~\cite{Lahaye2009,Bottcher2021,Chomaz2023}. The long-range and anisotropic dipolar interaction has enabled the observation of quantum droplets stabilized by beyond-mean-field fluctuations~\cite{Kadau2016,Schmitt2016,Ferrier-Barbut2016}, roton softening as a precursor to density instability~\cite{Petter2019,Hertkorn2021}, and supersolid phases in which crystalline order coexists with superfluidity~\cite{Tanzi2019,Bottcher2019,Chomaz2019,Guo2019,Tanzi2019a,Natale2019a,Sohmen2021}.  Together with advances in reduced geometries~\cite{Kao2021,Yang2024,Du2024,He2025}, correlated lattice systems~\cite{Aikawa2014b,Su2023a}, spin--orbital dynamics~\cite{Burdick2016,Matsui2025}, and synthetic-dimensional topology~\cite{Bouhiron2024}, these developments establish highly magnetic atoms as a versatile platform for exploring long-range-interacting quantum matter.

Introducing a second component changes the dipolar many-body problem qualitatively. Already in short-range interacting gases, binary mixtures have enabled interaction-controlled miscibility and coupled collective modes~\cite{Hall1998,Papp2008a,Tojo2010}, polaron physics~\cite{Schirotzek2009,Jorgensen2016}, and mediated interactions~\cite{DeSalvo2019,Edri2020}. In dipolar mixtures, long-range anisotropic interactions further couple these density and composition to interfacial structure, roton softening, and crystallization. Here and below, spin refers to the relative-density or composition channel, rather than the real magnetic spin of atoms. Theory predicts a broad hierarchy of phenomena, including roton-mediated phase separation~\cite{Wilson2012}, density, spin, and mixed roton instabilities~\cite{Lee2022}, composite droplets and coupled density-modulated states~\cite{Saito2009,Smith2021,Bisset2021,Bland2022,Halder2023}, as well as component-selective supersolidity and hybridized collective modes in binary supersolids~\cite{Li2022,Scheiermann2023,Kirkby2024,Scheiermann2025}. Dipolar mixtures therefore offer a route to quantum many-body phases in which spatial order can be shared, transferred, or separated between two coupled superfluids.




Experimentally, realizing such dipolar mixtures requires combining several demanding capabilities. The two components should be simultaneously condensed, strong spatial overlap, and subject to tunable intra- and interspecies interactions, while still allowing component-selective addressability and characterization. Recent progress includes strongly dipolar Er--Dy mixtures with calibrated interspecies Feshbach resonances~\cite{Trautmann2018,Durastante2020}, as well as spin mixtures of ultracold dysprosium with suppressed dipolar relaxation~\cite{Barral2024,Lecomte2025}. These experiments demonstrated key ingredients for multicomponent dipolar gases, but combining them in a stable and tunable binary dipolar Bose gas remains a central challenge.


Isotope mixtures of highly magnetic lanthanide atoms offer a combination of matched single-particle Hamiltonians and tunable interactions. Dysprosium is particularly well suited because its small isotope shifts and rich narrow-line spectrum enable simultaneous cooling, trapping, and isotope-selective addressability within an apparatus that remains essentially single-species-like. The nearly identical masses and optical polarizabilities of $^{162}$Dy and $^{164}$Dy provide naturally matched trapping potentials and strong spatial overlap, while dense intra- and interspecies Feshbach spectra allow magnetic tuning of the interactions. Together with controllable isotope loading, these features provide access to both interaction and composition control in a strongly dipolar Bose mixture.

Here we report the realization of a new quantum-degenerate dipolar mixture based on dysprosium isotopes, demonstrated with binary Bose--Einstein condensates (BECs) of $^{162}$Dy and $^{164}$Dy in a single-species-like experimental architecture. Starting from isotope-mixed laser cooling, we directly evaporate the mixture at each target magnetic field, yielding binary condensates with near-equilibrium density profiles set by the corresponding interaction balance. Isotope-resolved in-situ imaging reveals a magnetic-field-dependent reconfiguration from overlapped miscible profiles to core--shell-like, side-by-side, and exchanged core--shell-like immiscible geometries. At fixed magnetic field, population imbalance reshapes the phase-separated structure, providing a complementary control knob. These results establish dysprosium isotope mixtures as a compact and versatile platform for controlling miscibility, interfaces, and spatial order in strongly dipolar quantum fluids.



\begin{figure}[tb!]
	\includegraphics[width=0.9\columnwidth]{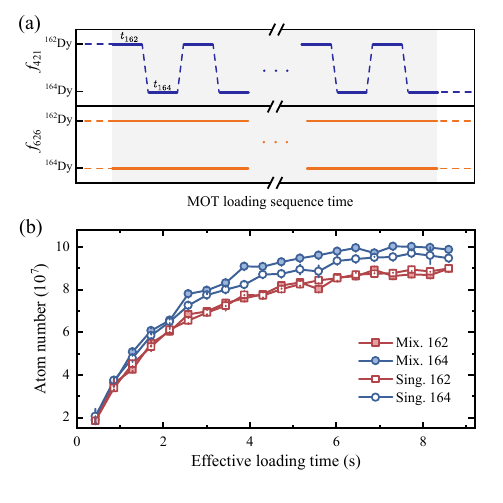}
	\caption{Isotope-mixed dysprosium MOT loading.
		(a) Timing sequence for dual-isotope MOT loading. During each loading shot marked as the shaded region, the 421~nm slowing and transverse-cooling light is switched between the $^{162}$Dy and $^{164}$Dy resonances for $n$ repeated cycles, with a dwell time $t_i$ per cycle and an effective loading time $n \times t_i$ on isotope $i$. The 626~nm narrow-line MOT light addresses both isotopes simultaneously. Dashed segments mark intervals in which the corresponding light is actively extinguished.
		(b) Atom number after the MOT-compression stage versus effective loading time for the isotope mixture and matched single-isotope references taken with the same 421~nm switching sequence. The close agreement of the loading curves indicates no measurable additional light-assisted loss in dual-isotope operation. All compressed clouds have temperatures of about $7.5~\mu\mathrm{K}$, and error bars denote the standard error of three measurements.}
	
	\label{fig:scheme}
\end{figure}

As illustrated in Fig.~\ref{fig:scheme}(a), simultaneous laser cooling and trapping of $^{162}$Dy and $^{164}$Dy are realized in a standard single-species dysprosium apparatus by frequency and timing control of the cooling light, without adding isotope-specific cooling or trapping beam paths. During each MOT-loading sequence, the 421~nm light for Zeeman slowing and transverse cooling is repeatedly switched between the two isotope resonances. In each cycle, the dwell time on isotope $i$ is $t_i$, giving an effective loading time $n\times t_i$ after $n$ cycles and allowing control of the relative loading of the two components. The sideband-engineered 626~nm narrow-line MOT light instead contains both isotope-resonant frequencies simultaneously. This scheme enables efficient dual-isotope loading into a spatially overlapped five-beam narrow-line magneto-optical trap (MOT)~\cite{Ilzhofer2017}, using essentially the same laser infrastructure and optical layout as a single-species experiment. The implementation is robust, combining seed-light modulation for 626-nm dual-frequency operation with reliable 421-nm switching and relocking over day-long operation.~\cite{SM,BMOTs}. Because the method relies primarily on spectral and temporal addressing rather than isotope-specific optical paths, it can be directly applied to other dysprosium isotope combinations, including Bose--Fermi and Fermi--Fermi mixtures, and can in principle be extended to mixtures with more than two isotopic components. More broadly, the same strategy should be applicable to related lanthanide species, and potentially to other atomic species with similar spectroscopic properties. Further details are given in the Supplementary Materials~\cite{SM}.

The loading dynamics of the isotope-mixed MOT, characterized by the atom number measured after the MOT-compression stage, are shown in Fig.~\ref{fig:scheme}(b) together with single-isotope reference measurements. The references use the same 421~nm switching sequence as the mixture experiment, but with the light extinguished during the dwell time assigned to the other isotope, and all compressed clouds have comparable temperatures of about $7.5~\mu$K. Under this matched comparison, the isotope-mixed MOT loading curves coincide with, and in some cases slightly exceed, their single-isotope counterparts. We therefore find no measurable additional light-assisted loss within our sensitivity. This robustness is unexpected, since near-resonant excitation in a mixed MOT could in principle open extra light-assisted collision channels. Together with a similar observation in dual-species Er-Dy MOTs~\cite{Ilzhofer2017}, our result suggests that suppressed light-assisted loss may be a broader feature of five-beam narrow-line lanthanide MOTs, although its microscopic origin remains to be clarified and lies beyond the scope of this work.

\begin{figure}[tb!]
	\includegraphics[width=0.9\columnwidth]{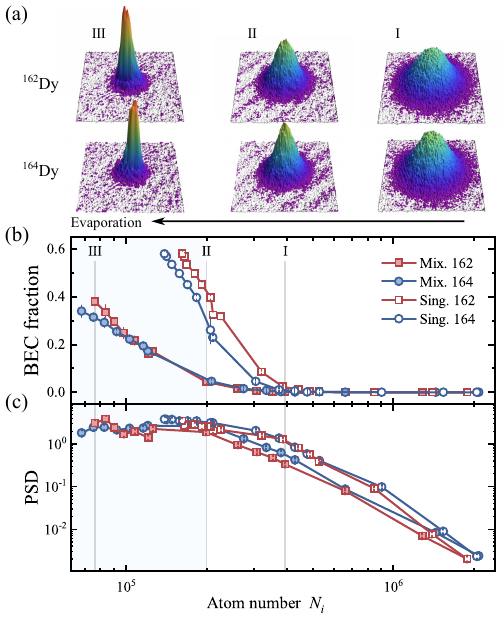}
	\caption{Evaporative cooling to binary BECs.
	(a) Time-of-flight density distributions of $^{162}$Dy and $^{164}$Dy after 35~ms expansion at three representative stages of the evaporation sequence at 9.62~G, showing the emergence of bimodal profiles in both isotopes.
	(b) Condensate fraction and (c) phase-space density versus atom number for the mixture and for single-isotope references prepared under comparable initial conditions. Here $N_i$ denotes the atom number of isotope $i$, and evaporation proceeds from right to left, and the vertical lines mark the three stages shown in (a). The shaded region makes the binary-condensates regime. Error bars denote the standard error of 10 independent measurements.}
	\label{fig:trajectory}
\end{figure}

After MOT compression, the mixture is directly transferred into a 1064~nm optical dipole trap and follows essentially the same cooling and trapping sequence as in single-species operation~\cite{SM}. We then perform forced evaporation on the mixture, which requires a favorable ratio of elastic to inelastic collisions in both intra- and interspecies channels to reach condensates. For strongly magnetic lanthanides, this ratio is highly magnetic-field dependent because of their dense Feshbach spectra. We therefore performed systematic intra- and interspecies Feshbach-loss spectroscopy of $^{162}$Dy and $^{164}$Dy over $0$--$32~\mathrm{G}$~\cite{FRs} to identify field regions suitable for evaporation and interaction tuning. Guided by this map, we use the 9.47--9.96~G range as an operating window for this work, where the mixture can be evaporatively cooled directly at the final magnetic field to produce binary condensates while the interactions remains magnetically tunable.
The relevant portion of the Feshbach spectrum is shown in the Supplementary Materials~\cite{SM}. Throughout the experiment, unless otherwise specified, the magnetic field is applied along the vertical $\hat{z}$ axis and stabilized to better than 1~mG.

We choose $9.62~\mathrm{G}$ as a representative field in this window to display the evaporation trajectory in Fig.~\ref{fig:trajectory}. Fig.~\ref{fig:trajectory}(a) shows representative absorption images after $35~\mathrm{ms}$ time-of-flight (TOF), and Fig.~\ref{fig:trajectory}(b)$\&$(c) compares the mixture evaporation with single-species references prepared under comparable initial conditions. Starting from approximately $2\times10^{6}$ atoms in each component at $T\simeq9~\mu$K, a $2.3~\mathrm{s}$ evaporation sequence produces binary condensates with $N_{\rm BEC}\simeq2.5\times10^{4}$ for each isotope, condensate fractions of about 0.4, and temperature near $90~\mathrm{nK}$ in a final cigar-shaped trap with $\omega_{x,y,z}=2\pi\times\left\lbrace 29(1),78(2),132(2)\right\rbrace$~Hz. Compared with the single-species references, the mixture condenses later and with reduced efficiency. From the pre-condensation evolution of phase-space density (PSD), we extract the evaporation efficiency $\gamma=-d\ln(\mathrm{PSD})/d\ln N$, obtaining $\gamma=3.1(1)$ and $3.2(1)$ for $^{162}$Dy and $^{164}$Dy in the mixture, compared with $3.9(2)$ and $3.5(2)$ for the single-species gases. Since the two isotopes experience nearly identical 1064~nm trapping potentials, no clear sympathetic-cooling hierarchy is expected, and the reduced efficiency is therefore attributed mainly to additional interspecies inelastic loss, consistent with Ref.~\cite{FRs}. The efficient evaporation is accompanied by long-lived binary condensates, as confirmed by the evolution measurements, shown in the Supplementary Materials~\cite{SM}, after a fast evaporation sequence, with sizable condensates remaining after $1~\mathrm{s}$ holding.




\begin{figure}[tb!]
	\includegraphics[width=0.9\columnwidth]{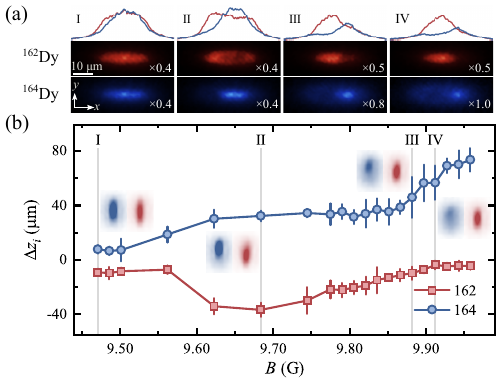}
	\caption{Magnetic-field-tuned spatial ordering.
	(a) Averaged in-situ density distributions of $^{162}$Dy and $^{164}$Dy at selected magnetic fields, obtained from more than 40 realizations. The line profiles above each column are one-dimensional densities along $x$, obtained by summing over the central rows of about 3~$\mu$m. Color scales are rescaled by the indicated factors for visibility, whereas line profiles remain unscaled for quantitative comparison. The images reveal a reconfiguration from core--shell-like to side-by-side and exchanged core--shell-like geometries. 
	(b) Vertical displacement of each condensed component after 35~ms TOF, measured relative to its thermal cloud. Larger values correspond to lower positions in real space. Vertical lines mark the four magnetic fields shown in (a), and the corresponding insets show TOF images averaged over ten shots. Error bars denote the standard error of the mean.}
	\label{fig:miscibility}
\end{figure}

Having established simultaneous condensation within the 9.47--9.96~G operating window, we probe magnetic-field-controlled miscibility and spatial ordering by evaporatively cooling the mixture directly at each final field. Although the evaporation efficiency is field dependent, this target-field preparation avoids post-condensation field ramp and allows the condensates to equilibrate under the interaction strengths being probed, yielding near-equilibrium density profiles close to the corresponding ground-state configurations. The central diagnostic is component-resolved imaging enabled by isotope-selective removal. Before each image, a resonant 626~nm push pulse removes the unwanted isotope with negligible perturbation to the remaining component~\cite{SM}. This allows us to perform isotope-resolved in-situ phase-contrast imaging along $\hat{z}$, which directly resolves the transverse density distribution in Fig.~\ref{fig:miscibility}(a), and side absorption imaging after $35~\mathrm{ms}$ TOF, which measures the vertical displacement of each condensate relative to its thermal background, $\Delta z_i=z_{i,\mathrm{BEC}}-z_{i,\mathrm{thermal}}$. Since each realization detects only one isotope, two-component observables are reconstructed from independent preparations under identical conditions.


As summarized in Fig.~\ref{fig:miscibility}, the in-situ and TOF probes provide consistent signatures of a magnetic-field-driven miscible--immiscible reconfiguration. On the low-field side, the in-situ images show strongly overlapped condensates, identifying a miscible binary BEC. The corresponding TOF displacement is much smaller than in the immiscible regime, but remains finite. This can be attributed, at least in part, to long-time expansion, which can greatly amplify a residual differential trap-center offset below $0.2~\mu$m arising from the isotope mass difference, with possible additional contributions from interaction-dependent dynamics during release. We therefore characterize miscibility from the correlated evolution of the in-situ density profiles and the TOF displacement. With increasing magnetic field, both probes reveal the emergence of phase separation and the development of large relative displacements.

Within the immiscible regime, the in-situ images reveal a field-driven reconfiguration of the condensate interface. As the magnetic field is increased, the density distribution evolves from a core--shell--like configuration with $^{164}$Dy occupying the central high-density region and $^{162}$Dy distributed outside, to a side-by-side geometry, and then toward an exchanged core--shell-like structure with $^{162}$Dy moves toward the center and $^{164}$Dy is displaced outward. This real-space reconfiguration is accompanied by large TOF displacements, with $^{162}$Dy above and $^{164}$Dy below the thermal backgrounds throughout the immiscible regime. Their magnitudes depend on the condensate population balance, with the more populated component generally showing a smaller $|\Delta z_i|$, consistent with an approximately conserved two-component center of mass.

The observed reconfiguration reflects the magnetic tuning of the interactions by nearby Feshbach resonances. In the 9.47--9.96~G range, the $^{162}$Dy intraspecies scattering length $a_{11}$ varies only weakly, whereas the $^{164}$Dy intraspecies scattering length $a_{22}$ is strongly modified by nearby resonances around 9.525~G, 9.995~G, 10.025~G, and 10.126~G. The interspecies scattering length $a_{12}$ also increases toward the broad interspecies resonance near 10.6~G and remains large enough to support phase separation~\cite{FRs}. Because the condensate populations are not balanced in this data set, the observed spatial order is governed not by the scattering lengths alone, but by density-dependent interaction energies, interspecies repulsion, and the kinetic-energy cost of the interface. At lower fields, the stronger $^{162}$Dy self-repulsion favors its occupation of the lower-density outer region. As $a_{22}$ increases, the growing $^{164}$Dy self-interaction drives the system toward a side-by-side interface and eventually toward the exchanged core--shell-like geometry.

\begin{figure}[tb!]
	\includegraphics[width=0.9\columnwidth]{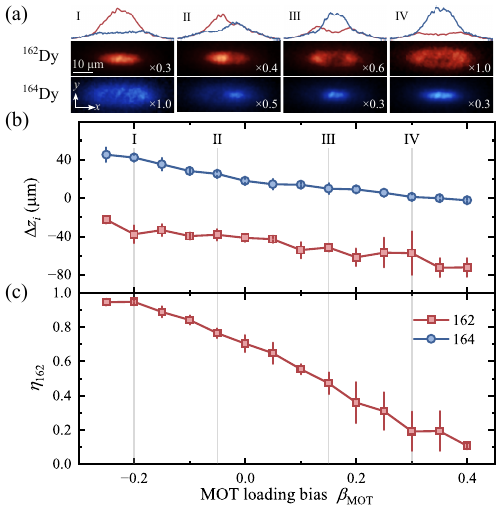}
	\caption{Population-imbalance control of binary condensates.
	(a) In-situ density distributions of $^{162}$Dy and $^{164}$Dy condensates for selected condensate number ratio at $9.88~{\rm G}$, plotted following the convention of Fig.~\ref{fig:miscibility}(a). The population imbalance is tuned through the isotope loading bias in the MOT.
	(b) Relative vertical displacement $\Delta z_i$ and (c) condensate number ratio $\eta_{162}$ versus MOT loading bias $\beta_{\mathrm{MOT}}$. Vertical lines mark the four conditions shown in (a). The line III at $\beta_{\mathrm{MOT}}=0.15$ indicates a near-balanced case with $\eta_{162}=0.47(6)$. Error bars denote standard errors.}
	\label{fig:ratio}
\end{figure}

To further probe the immiscible phase, we tune the relative condensate populations at a fixed magnetic field of 9.88~G by varying the isotope dwell times in the MOT loading sequence~\cite{SM}. The loading bias $\beta_{\mathrm{MOT}}=(t_{164}-t_{162})/(t_{162}+t_{164})$ systematically controls the final condensate number ratio $\eta_{162}=N_{162}^{\rm BEC}/(N_{162}^{\rm BEC}+N_{164}^{\rm BEC})$, providing a composition-control knob at fixed interactions.

For comparable condensate populations in matched traps, the immiscible geometry provides a qualitative diagnostic of the relative self-interactions. In the near-balanced condition at $9.88~\mathrm{G}$, III in Fig.~\ref{fig:ratio}(a), $^{162}$Dy occupies the lower-density outer region, whereas $^{164}$Dy remains closer to the central high-density region, indicating that $^{162}$Dy is more strongly self-repulsive and hence $a_{11}>a_{22}$ at this field. Together with the Feshbach spectra discussed above, this suggests that $a_{11}>a_{22}$ also holds across the 9.53--9.88~G range. Thus, the spatial order provides a useful indicator of the hierarchy of their interactions.


Starting from this hierarchy, changing $\beta_{\mathrm{MOT}}$ reshapes the phase-separated state through population imbalance. As $\beta_{\mathrm{MOT}}$ is increased, $^{162}$Dy evolves from the majority to the minority component, and the in-situ density profiles change from a $^{162}$Dy-rich configuration, through more balanced side-by-side geometries, to a $^{164}$Dy-rich configuration in which the depleted $^{162}$Dy component is displaced toward the edge. The TOF displacements in Fig.~\ref{fig:ratio}(b) provide a complementary vertical signature, with $|\Delta z_i|$ decreasing as the corresponding condensate fraction increases.
Together, the in-situ profiles and TOF positions demonstrate that population imbalance tunes the spatial partitioning of the phase-separated condensates at fixed interactions, through the density dependence of the mean-field interaction energies and the kinetic cost of forming an interface.

In summary, we realize a new quantum-degenerate dipolar mixture and establish a dysprosium-isotope platform for strongly dipolar quantum mixtures, demonstrated here with binary condensates of $^{162}$Dy and $^{164}$Dy. The mixture is produced in a single-species-like apparatus, while combining nearly matched single-particle Hamiltonians and tunable intra- and interspecies interactions. Guided by the Feshbach spectrum, we evaporatively cool the mixture directly at target magnetic fields and resolve the resulting spatial order with component-selective imaging. Interaction tuning drives a miscible--immiscible reconfiguration in which the two condensates do not merely separate, but reshape their interface from core--shell-like to side-by-side and exchanged core--shell-like geometries. Population imbalance provides a second control axis, tuning the phase-separated geometry at fixed scattering lengths through the density-dependent mean-field energy balance. Together, these capabilities establish dysprosium isotope mixtures as a compact and extensible platform for strongly dipolar two-component quantum fluids, opening routes toward binary supersolidity, coupled density--spin crystallization, and impurity physics in strongly imbalanced mixtures.



\begin{acknowledgements}
	
	We thank Yongchang Zhang and Kuitian Xi for fruitful discussion. This work is supported by the National Key R$\&$D Program of China (No.~2022YFA1405800), Innovation Program for Quantum Science and Technology (No.~2024ZD0300600), National Natural Science Foundation of China (No.~12474263), Guangdong Basic and Applied Basic Research Foundation (No.~2023B0303000011), Guangdong Provincial Quantum Science Strategic Initiative (No.~GDZX2304006, GDZX2303002, GDZX2403002), Shenzhen Science and Technology Program (No.~KQTD20240729102026004), Stable Support Plan Program of Shenzhen Natural Science Fund (No.~20220815001356001), and Guangdong Provincial Key Laboratory of Advanced Thermoelectric Materials and Device Physics (No.~2024B1212010001).
	
\end{acknowledgements}

\bibliographystyle{apsrev4-2}
\bibliography{refs_all_clean}

\setcounter{figure}{0}
\renewcommand{\thefigure}{S\arabic{figure}}

\setcounter{table}{0}
\renewcommand{\thetable}{S\arabic{table}}

\setcounter{equation}{0}
\renewcommand{\theequation}{S\arabic{equation}}

\clearpage
\onecolumngrid

\section{Supplementary Material}

\subsection{Experimental setup and dual-isotope laser cooling}

The experiment is performed on a newly constructed dysprosium apparatus that retains the standard laser-cooling and optical-trapping architecture used for single-species dysprosium experiments. The sequence consists of Zeeman slowing, transverse cooling, narrow-line magneto-optical trapping, transfer into crossed optical dipole traps (ODTs), and forced evaporation. Dual-isotope operation is implemented by frequency addressing the cooling and imaging transitions of $^{162}$Dy and $^{164}$Dy, without adding a separate slowing, cooling, or trapping beam path.

For the broad 421~nm transition used for Zeeman slowing, transverse cooling and imaging, a single laser source, with its frequency referenced to a cavity, sequentially addresses the $^{162}$Dy and $^{164}$Dy transitions by switching the radio-frequency drive applied to a fiber electro-optic modulator in the locking path. The dwell time on isotope $i$ during each switching cycle is denoted by $t_i$. After each frequency jump, the lock is reproducibly reacquired at the new operating point within $70~\mathrm{ms}$, during which the 421~nm light is actively switched off to suppress uncontrolled scattering and transient effects. With optimized locking and switching parameters, this alternating frequency-switching sequence operates reliably during day-long experiments~\cite{BMOTs}. By alternating the 421~nm light frequency between the two isotope resonances, this scheme provides stable Zeeman slowing and transverse cooling for both isotope beams. The dwell time at each isotope resonance sets the effective loading time of the corresponding isotope, providing a simple and reproducible control of the population imbalance while leaving the subsequent cooling and trapping sequence unchanged.

For the narrow-line 626~nm transition used for magneto-optical trapping, the light contains two simultaneous frequency components, each resonant with one isotope. The two isotopes are therefore cooled and confined simultaneously in the same MOT volume. The 626~nm light is generated by sum-frequency generation of 1550~nm and 1050~nm lasers, allowing the 626~nm spectrum to be engineered at the fundamental wavelengths. For dual-isotope operation, we modulate the 1550~nm seed laser with a fiber electro-optic modulator. This seed-laser-level modulation provides a flexible and robust way to generate controlled spectral components at 626~nm, and can in principle be extended to programmable multi-frequency spectra with more complex amplitude and frequency structures~\cite{Holland2021}.

The frequency-switched 421~nm cooling light and the dual-frequency 626~nm MOT light are combined in a synchronized loading sequence, as illustrated in Fig.~\ref{fig:scheme}. During Zeeman slowing and transverse cooling, the 421~nm light alternately addresses the two isotopes, whereas in the MOT region the 626~nm light acts on both isotopes simultaneously. This combination enables the formation of an isotope-mixed narrow-line MOT within an otherwise single-species apparatus. Importantly, since the scheme relies on frequency addressing rather than isotope-specific optical paths, the same strategy can be adapted to other dysprosium isotope combinations, including Bose--Fermi and Fermi--Fermi mixtures, and more broadly to atomic species with similarly suitable spectroscopic structures.

\subsection{Optical trapping and evaporation of the isotope mixture}

After a $100~\mathrm{ms}$ MOT-compression stage, the isotope mixture is directly loaded into a single-beam 1064~nm optical dipole trap using the standard single-species Dy sequence. The trap is formed by a horizontal beam propagating along $\hat y$, with a vertical waist of about $30~\mu\mathrm{m}$. Transverse confinement along $\hat x$ is generated by acousto-optic deflection (AOD), producing an elongated time-averaged potential whose aspect ratio is optimized during loading and evaporation. In-trap narrow-line cooling on the 626~nm transition is then applied in the single-beam ODT, increasing the phase-space density by more than an order of magnitude. The gas is subsequently transferred into a crossed ODT by adding a second horizontal beam along $\hat x$, orthogonal to the first beam, with a waist of about $50~\mu\mathrm{m}$.

\begin{figure}[tb!]
	\includegraphics[width=0.45\columnwidth]{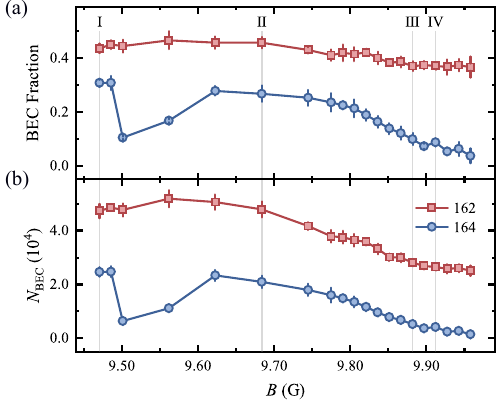}
	\caption{Binary condensate condition across the magnetic-field scan for Fig.~\ref{fig:miscibility}.
		(a) BEC fraction and (b) condensed atom number $N_{\rm BEC}$ of $^{162}$Dy and $^{164}$Dy after direct evaporation at different final magnetic fields. The data are extracted from the same TOF images as in Fig.~\ref{fig:miscibility}(b). Both isotopes remain condensed over the explored field range. Error bars denote the standard error of the mean.}
	
	\label{fig:fig3_sm}
\end{figure}

For the benchmark evaporation trajectory shown in Fig.~\ref{fig:trajectory}, the magnetic field is held fixed at $9.62~\mathrm{G}$. The sequence starts with approximately $2\times10^{6}$ atoms in each component at $T\simeq9~\mu\mathrm{K}$, with initial trap frequencies $\omega_{x,y,z}=2\pi\times\left\lbrace 56(3),588(37),826(36)\right\rbrace~\mathrm{Hz}$. During the $2.15~\mathrm{s}$ forced-evaporation stage, the ODT power and AOD-scanned trap geometry are optimized, reaching final trap frequencies of $\omega_{x,y,z}=2\pi\times\left\lbrace 29(1),78(2),132(2)\right\rbrace~\mathrm{Hz}$. After reaching the final trap, the atoms are held for a variable time. The final condition quoted in the main text is measured after a $120~\mathrm{ms}$ hold in the final trap, giving typical condensed atom numbers of $N_{162}^{\rm BEC}\simeq N_{164}^{\rm BEC}\simeq2.5\times10^{4}$, condensate fractions of about 0.4, and temperatures near $90~\mathrm{nK}$. The last six points in Fig.~\ref{fig:trajectory} correspond to different hold times in the final trap.


The evaporation efficiency is extracted from the thermal part of the trajectory before the appearance of a condensate. We fit the phase-space density as a function of atom number and define the evaporation efficiency parameter $\gamma=-d\ln(\mathrm{PSD})/d\ln N$. The mixture gives $\gamma=3.1(1)$ for $^{162}$Dy and $3.2(1)$ for $^{164}$Dy, compared with $3.9(2)$ and $3.5(2)$ for the corresponding single-species references. These values show that both components evaporate efficiently in the mixture. Compared with single-species references taken with the same mixture-optimized evaporation sequence, the modest reduction in $\gamma$ is consistent with additional loss and thermalization dynamics introduced by the second component.

For the spatial-ordering and population-imbalance measurements in Figs.~3 and 4, we instead use a two-step field sequence. The mixture is first evaporated at $9.62~\mathrm{G}$ to just above the condensation threshold, after which the magnetic field is rapidly ramped to the target value. The final evaporation is then carried out at this field to prepare the desired many-body state. Unless otherwise specified, the gas is held for an additional $90~\mathrm{ms}$ in the final trap before detection.

An important aspect of this preparation is that the condensate forms during evaporation at the target interaction strength, rather than being produced by a diabatic interaction ramp from a preformed condensate. Excess energy is therefore continuously removed as the system approaches the final state, reducing ramp-induced excitations and possible metastability. This direct preparation allows the binary condensate to relax toward its equilibrium spatial configuration at the target interaction strength.

The corresponding sample parameters for Fig.~3 are shown in Fig.~\ref*{fig:fig3_sm}, including the condensed atom number and condensate fraction of each isotope as functions of magnetic field. The temperature remains in the range of $90$--$110~\mathrm{nK}$ for all magnetic fields.

\begin{figure}[tb!]
	\includegraphics[width=0.45\columnwidth]{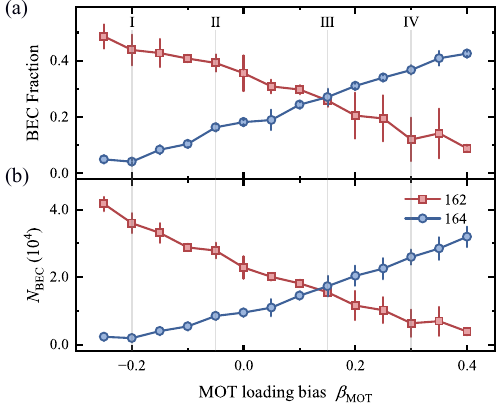}
	\caption{Binary condensate condition as a function of MOT loading bias.
	(a) BEC fraction and (b) condensed atom number $N_{\rm BEC}$ of $^{162}$Dy and $^{164}$Dy after direct evaporation at different MOT-loading biases. The data are extracted from the same TOF images as in Fig.~\ref{fig:ratio}(b). The MOT-loading bias continuously tunes the condensate population from $^{162}$Dy-dominated to $^{164}$Dy-dominated samples. Error bars denote the standard error of the mean.}
	
	\label{fig:fig4_sm}
\end{figure}

For Fig.~4, Fig.~\ref*{fig:fig4_sm} summarizes the condensed atom numbers and condensate fractions as functions of the MOT loading bias. The temperature remains in the range of $80$--$100~\mathrm{nK}$, with a slight decrease toward the $^{164}$Dy-rich side.

\subsection{Component-resolved detection and analysis}

Component-resolved detection is performed with 421~nm probe light, using either in-situ phase-contrast imaging or resonant absorption imaging after TOF expansion. In-situ density measurements are taken along the vertical direction $\hat z$, and therefore probe the column-density distribution in the $x$-$y$ plane with a resolution of about $1.5~\mu\mathrm{m}$. For phase-contrast imaging, the probe is detuned by $-15.3~\Gamma$ from the transition of the imaged isotope, where $\Gamma$ denotes the natural linewidth of the 421~nm transition. This imaging geometry is used to characterize the real-space density distribution and spatial ordering of the two condensates.

Atom numbers, condensate fractions, temperatures, and vertical center-of-mass positions are extracted from resonant absorption images taken after TOF from a horizontal side-view direction, tilted by about $22.5^\circ$ relative to the $\hat y$ axis. Unless otherwise specified, the side-view images are recorded after $35~\mathrm{ms}$ of expansion. The density profiles are fitted with Gaussian or bimodal distributions, depending on the evaporation stage, to extract the thermal and condensed components and their vertical positions.

In each experimental realization, only one isotope is imaged. The other isotope is selectively removed from the imaging region by a resonant 626~nm light pulse propagating along the same direction of the side-imaging beam. The pulse duration, normally between 100~$\mu s$ to 500~$\mu s$, is chosen such that the removed component is displaced by more than three times the cloud size, while the imaged component remains essentially unaffected. The experiment alternates between imaging $^{162}$Dy and $^{164}$Dy, and component-resolved observables are obtained from repeated realizations under identical preparation conditions.

\subsection{Feshbach loss spectroscopy}

\begin{figure*}[t!]
	\centering
	\includegraphics[width=0.8\columnwidth]{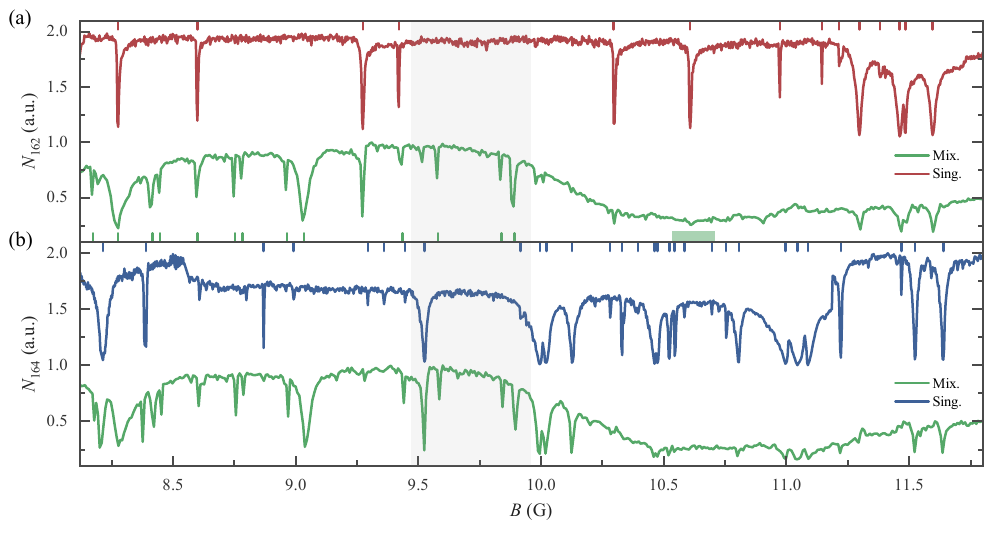}
	\caption{Feshbach loss spectra near the dual-condensate working fields.
		Remaining atom number of $^{162}$Dy (a) and $^{164}$Dy (b) for single-species samples and isotope mixtures. The main-text working window, $9.47$--$9.96~\mathrm{G}$ marked as shaded region, lies on the low-field side of a broad inters-pecies loss feature near $10.6~\mathrm{G}$. Colored ticks on the horizontal axes mark identified resonances, with $^{162}$Dy, $^{164}$Dy, and interspecies features shown in red, blue, and green, respectively.}
	\label{fig:FRs_sm}
\end{figure*}

Feshbach-loss spectroscopy is performed for both single-species samples and isotope mixtures over $0$--$32~\mathrm{G}$~\cite{FRs}. Figure~\ref{fig:FRs_sm} shows the $8$--$12~\mathrm{G}$ region relevant to this work, covering the main evaporation and interaction-tuning fields between $9.47$ and $9.96~\mathrm{G}$. The magnetic field is actively stabilized to better than $1~\mathrm{mG}$ during these measurements.

For the single-species measurements, loss spectra are recorded during forced evaporation. A single-isotope gas is first cooled at a fixed preparation field to approximately $1~\mu\mathrm{K}$. The magnetic field is then ramped to the target value, and evaporation is continued for $180~\mathrm{ms}$ to a final temperature of approximately $0.5~\mu\mathrm{K}$. The remaining atom number is measured after TOF. The field step is approximately $3~\mathrm{mG}$.

For the isotope-mixture measurements, loss spectra are obtained by hold-time spectroscopy. The mixture is first evaporatively cooled at a fixed preparation field and then rapidly ramped to the target magnetic field. After a hold time of $200~\mathrm{ms}$, the remaining atom numbers of both isotopes are measured. In the main working region of $9.47$--$9.96~\mathrm{G}$, the spectra are taken at approximately $1~\mu\mathrm{K}$ with a field step of about $3~\mathrm{mG}$.

The resolved loss features provide a practical map of the resonant structure relevant to evaporation and interaction tuning. In the working field range, $^{162}$Dy shows only one narrow loss feature near $9.42~\mathrm{G}$, indicating that its intraspecies interaction is expected to vary only weakly away from this resonance. In contrast, $^{164}$Dy exhibits several narrow resonances near $9.525~\mathrm{G}$, $9.995~\mathrm{G}$, $10.025~\mathrm{G}$, and $10.126~\mathrm{G}$, together with a weaker feature near $9.916~\mathrm{G}$. The mixture spectra additionally show a very broad loss feature centered near $10.6~\mathrm{G}$, which is absent in the corresponding single-species spectra and is therefore assigned to an interspecies Feshbach resonance.

\subsection{Post-evaporation formation and lifetime of the binary condensates}

\begin{figure*}[t!]
	\centering
	\includegraphics[width=0.45\columnwidth]{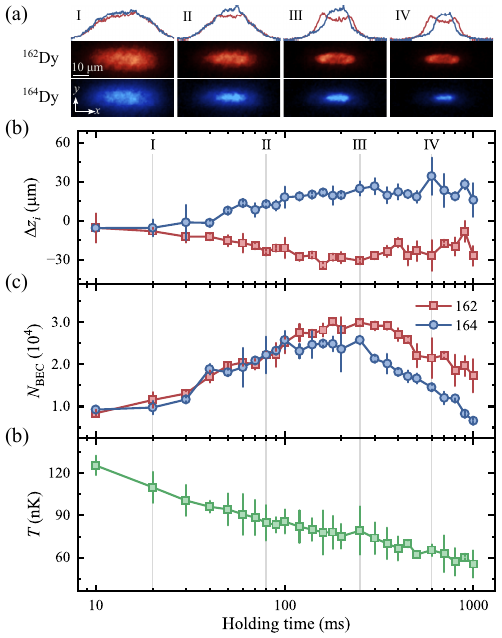}
	\caption{Post-evaporation evolution of the binary condensates.
		(a) Component-resolved in-situ density distributions of $^{162}$Dy and $^{164}$Dy at selected hold times after the fast evaporation sequence, plotted following the convention of Fig.~\ref{fig:miscibility}(a) with a slightly larger field of view.
		(b) Relative vertical displacement $\Delta z_i$, (c) condensed atom number, and (d) averaged temperature of the two isotopes versus hold time. The temperatures of the two isotopes agree within the measurement uncertainty throughout the evolution. Error bars denote standard errors
		}
	\label{fig:dynamics_sm}
\end{figure*}

To characterize the formation and stability of the binary condensates, we perform an additional evolution measurement at $9.62~\mathrm{G}$ after a shortened evaporation sequence. In this measurement, the forced-evaporation time is reduced from $2.15~\mathrm{s}$ to $1.25~\mathrm{s}$, such that condensates only begin to appear at the end of the evaporation stage. The mixture is then held in the final trap, and its subsequent evolution is monitored using component-resolved in-situ imaging and side-view absorption imaging after TOF expansion, as shown in Fig.~\ref{fig:dynamics_sm}.

The condensed atom numbers initially increase during the hold time, indicating continued cooling and condensate growth after the shortened evaporation sequence, and then decrease at longer times due to atom loss. Nevertheless, sizable condensates remain visible after $1~\mathrm{s}$ of hold time, providing a lifetime benchmark for the binary condensates. The temperature decreases throughout the evolution, from about $120~\mathrm{nK}$ to about $60~\mathrm{nK}$.

The spatial ordering extracted from the in-situ images also evolves during the hold. At early times, when the thermal fraction is large, the thermal background partially masks the condensate density distribution, and the two components appear to have substantial overlap in the in-situ images. As the condensate fraction increases, the immiscible density pattern becomes well resolved and persists after $1~\mathrm{s}$ of hold time. The TOF side-view measurements provide a consistent signature, with the relative vertical displacement becoming clearly resolved once the condensate fraction is sufficiently high.

This measurement captures the post-evaporation evolution after a shortened evaporation sequence, with sizable binary condensates and well-resolved immiscible spatial order persisting after $1~\mathrm{s}$ in the final trap.

\clearpage

\end{document}